\documentclass[aip, twocolumn]{revtex4-1}
\usepackage{graphicx}
\usepackage{caption}
\usepackage{dcolumn}
\usepackage{bm}
\usepackage[utf8]{inputenc}
\usepackage[T1]{fontenc}
\usepackage{mathptmx}
\usepackage{etoolbox}
\usepackage{subfig}
\usepackage{braket}
\usepackage{hyperref}
\usepackage[hyphenbreaks]{breakurl}

\makeatletter
\def\@email#1#2{%
 \endgroup
 \patchcmd{\titleblock@produce}
  {\frontmatter@RRAPformat}
  {\frontmatter@RRAPformat{\produce@RRAP{*#1\href{mailto:#2}{#2}}}\frontmatter@RRAPformat}
  {}{}
}%
\makeatother

\begin{document}

\title{Fabrication and Characterization of Impedance-transformed Josephson Parametric Amplifier} 


\author{Zhengyang Mei}
 \affiliation{Beijing National Laboratory for Condensed Matter Physics, Institute of Physics, Chinese Academy of Sciences, Beijing 100190, China}
 \affiliation{School of Physical Sciences, University of Chinese Academy of Sciences, Beijing 100049, China}

\author{Xiaohui Song}
 \affiliation{Beijing National Laboratory for Condensed Matter Physics, Institute of Physics, Chinese Academy of Sciences, Beijing 100190, China}
\affiliation{Beijing Key Laboratory of Fault-Tolerant Quantum Computing, Beijing Academy of Quantum Information Sciences, Beijing 100193, China}
 \affiliation{Hefei National Laboratory, Hefei 230088, China}

\author{Xueyi Guo}
 \affiliation{Beijing Key Laboratory of Fault-Tolerant Quantum Computing, Beijing Academy of Quantum Information Sciences, Beijing 100193, China}
 
\author{Xiang Li}
 \affiliation{Beijing National Laboratory for Condensed Matter Physics, Institute of Physics, Chinese Academy of Sciences, Beijing 100190, China}
 \affiliation{School of Physical Sciences, University of Chinese Academy of Sciences, Beijing 100049, China}
 
\author{Yunhao Shi}
 \affiliation{Beijing National Laboratory for Condensed Matter Physics, Institute of Physics, Chinese Academy of Sciences, Beijing 100190, China}

\author{Guihan Liang}
 \affiliation{Beijing National Laboratory for Condensed Matter Physics, Institute of Physics, Chinese Academy of Sciences, Beijing 100190, China}
 \affiliation{School of Physical Sciences, University of Chinese Academy of Sciences, Beijing 100049, China}

\author{Chenglin Deng}
 \affiliation{Beijing National Laboratory for Condensed Matter Physics, Institute of Physics, Chinese Academy of Sciences, Beijing 100190, China}
 \affiliation{School of Physical Sciences, University of Chinese Academy of Sciences, Beijing 100049, China}

\author{Li Li}
 \affiliation{Beijing National Laboratory for Condensed Matter Physics, Institute of Physics, Chinese Academy of Sciences, Beijing 100190, China}
 \affiliation{School of Physical Sciences, University of Chinese Academy of Sciences, Beijing 100049, China}

\author{Yang He}
 \affiliation{Beijing National Laboratory for Condensed Matter Physics, Institute of Physics, Chinese Academy of Sciences, Beijing 100190, China}
 \affiliation{School of Physical Sciences, University of Chinese Academy of Sciences, Beijing 100049, China}

\author{Dongning Zheng}
 \affiliation{Beijing National Laboratory for Condensed Matter Physics, Institute of Physics, Chinese Academy of Sciences, Beijing 100190, China}
 \affiliation{School of Physical Sciences, University of Chinese Academy of Sciences, Beijing 100049, China}
\affiliation{Hefei National Laboratory, Hefei 230088, China}
 \affiliation{Songshan Lake Materials Laboratory, Dongguan 523781, China}

\author{Kai Xu}
   \email{kaixu@iphy.ac.cn}
 \affiliation{Beijing National Laboratory for Condensed Matter Physics, Institute of Physics, Chinese Academy of Sciences, Beijing 100190, China}
 \affiliation{School of Physical Sciences, University of Chinese Academy of Sciences, Beijing 100049, China}
 \affiliation{Beijing Key Laboratory of Fault-Tolerant Quantum Computing, Beijing Academy of Quantum Information Sciences, Beijing 100193, China}
 \affiliation{Hefei National Laboratory, Hefei 230088, China}
 \affiliation{Songshan Lake Materials Laboratory, Dongguan 523781, China}

\author{Heng Fan}
  \email{hfan@iphy.ac.cn}
 \affiliation{Beijing National Laboratory for Condensed Matter Physics, Institute of Physics, Chinese Academy of Sciences, Beijing 100190, China}
 \affiliation{School of Physical Sciences, University of Chinese Academy of Sciences, Beijing 100049, China}
 \affiliation{Beijing Key Laboratory of Fault-Tolerant Quantum Computing, Beijing Academy of Quantum Information Sciences, Beijing 100193, China}
 \affiliation{Hefei National Laboratory, Hefei 230088, China}
 \affiliation{Songshan Lake Materials Laboratory, Dongguan 523781, China}

 \author{Zhongcheng Xiang}
   \email{zcxiang@iphy.ac.cn}
 \affiliation{Beijing National Laboratory for Condensed Matter Physics, Institute of Physics, Chinese Academy of Sciences, Beijing 100190, China}
 \affiliation{School of Physical Sciences, University of Chinese Academy of Sciences, Beijing 100049, China}
 \affiliation{Hefei National Laboratory, Hefei 230088, China}



\date{\today}

\begin{abstract}
In this paper, we introduce a method of using a double-layer resist lift-off process to prepare the capacitor dielectric layer for fabricating impedance-engineered Josephson parametric amplifiers (IMPAs). Compared with traditional techniques, this method enhances fabrication success rate, accelerates production. The IMPA we made experimentally achieves an instantaneous bandwidth over 950 (600) MHz with a gain exceeding 10 (14) dB, along with saturation input power of -115 dBm and near quantum-limited noise. We demonstrate the negligible backaction from the IMPA on superconducting qubits, resulting in no significant degradation of the relaxation time and coherence time of the qubits. The IMPA improves the signal-to-noise ratio from 1.69 to 14.56 and enables the amplification chain to achieve a high quantum efficiency with $\eta \approx 0.26$, making it a critical necessity for large-scale quantum computation.

\end{abstract}


\maketitle 


Quantum-limited\cite{Caves1982} Josephson parametric amplifiers (JPAs) play a key role in experiments for detection of single-photon-level microwave signals in the field of circuit quantum electrodynamics. Featured applications include single-shot readout,\cite{Lin2013,Abdo2011,Walter2017} feedback control of superconducting qubits,\cite{Vijay2012,Ristè2013} and the generation of squeezed quantum states.\cite{Yurke1987,Castellanos-Beltran2008} Despite extensive research on Josephson parametric amplifiers over the past decades,\cite{Mooij1999,You2003,Koch2007,Yamamoto2008,Reed2010,Hatridge2011,Lucero2012,Barends2013,Zhong2013} their limited bandwidth and dynamic range have constrained their applicability in the rapidly advancing field of multi-qubit architectures and the realization of fault-tolerant quantum computation.\cite{Chen2012,Fowler2012,Barends2014}

\begin{figure*}[t]
    \centering
    \subfloat[\label{fig1a}]{\includegraphics[width=0.46\linewidth, trim=11.5cm 4.8cm 4cm 0cm, clip]{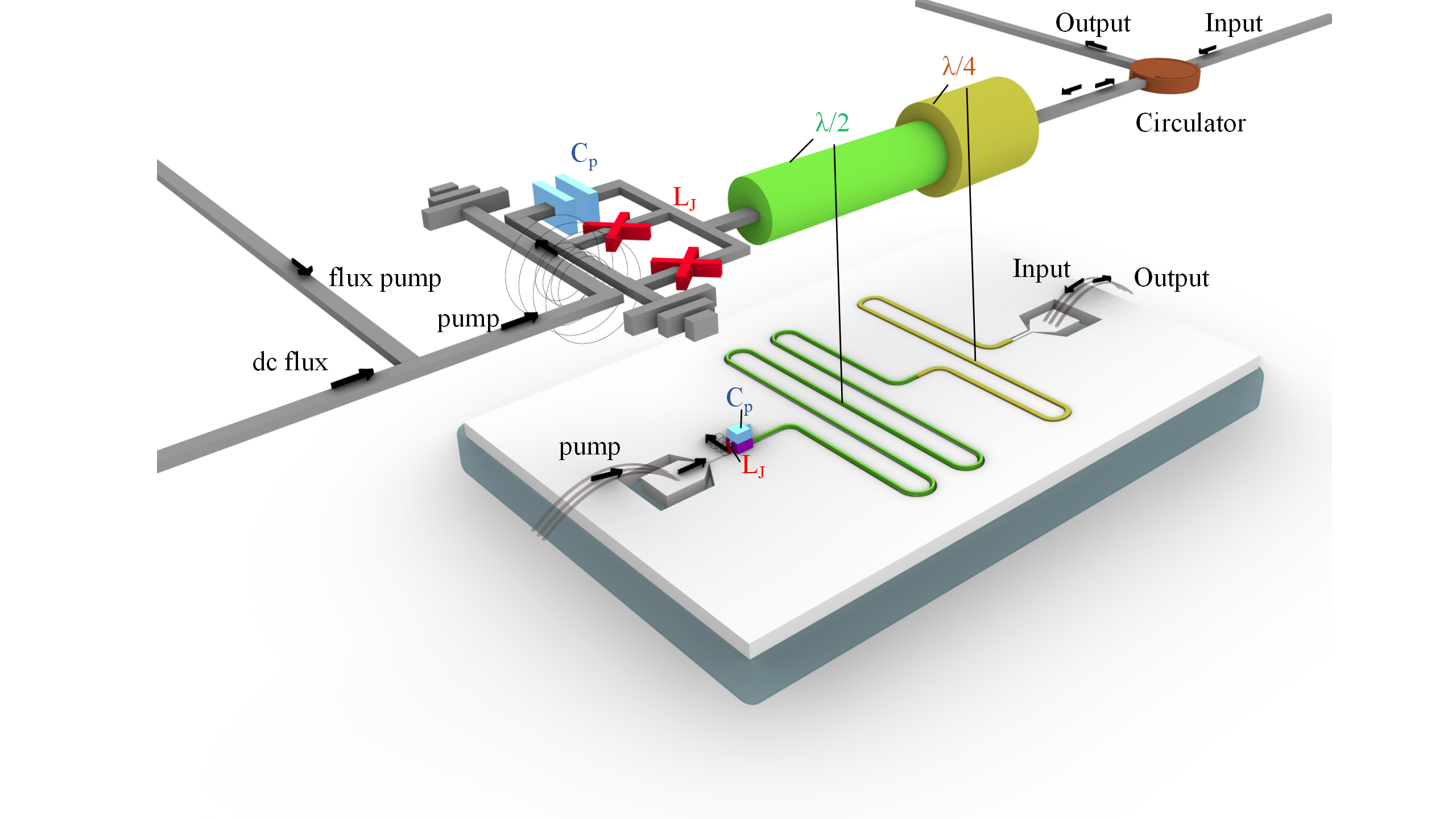}}
    \hfill
    \subfloat[\label{fig1b}]{\includegraphics[width=0.522\linewidth]{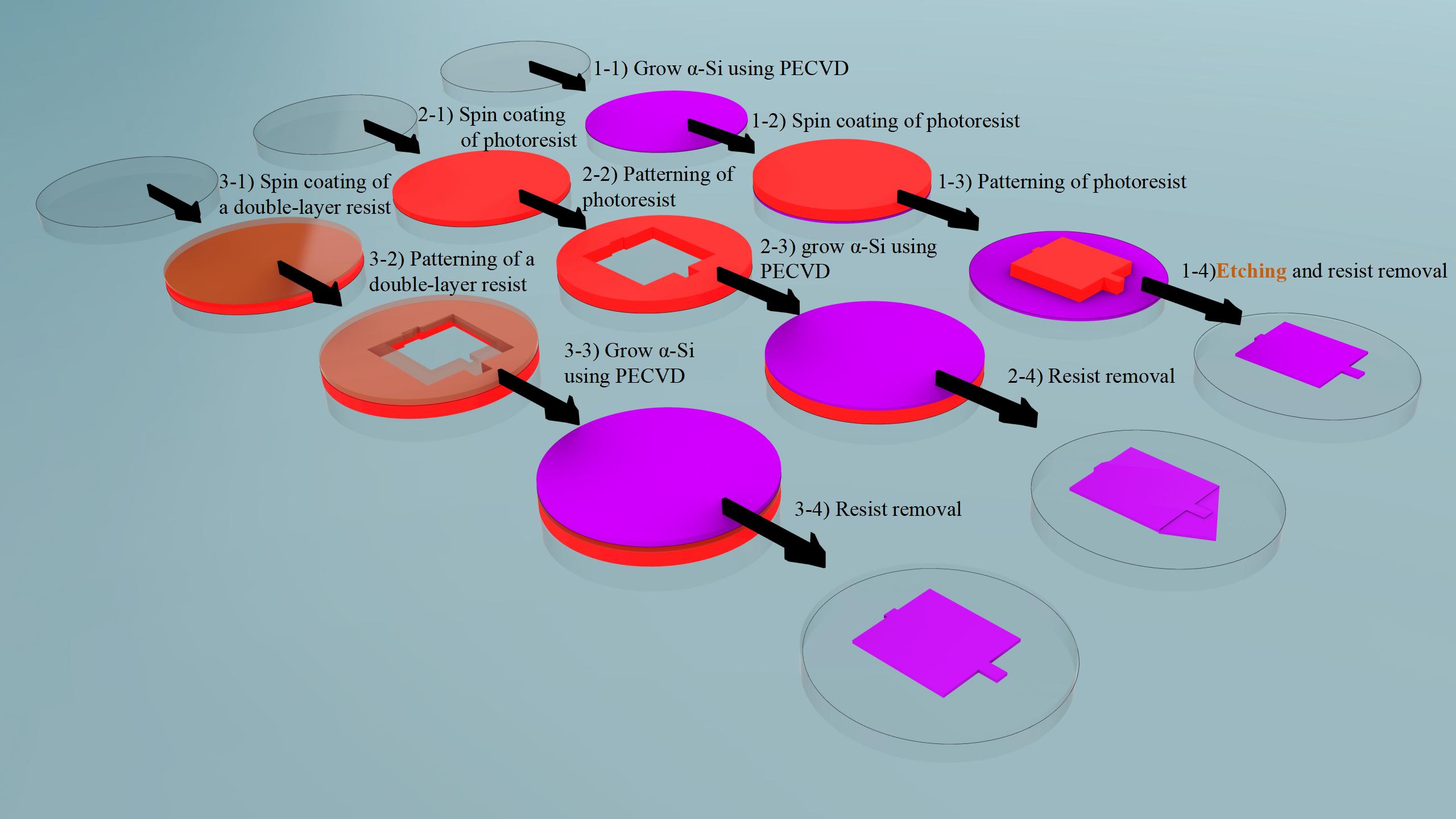}}
    \captionsetup{justification=raggedright,singlelinecheck=false}
   \caption{The device and different fabrication methods of its dielectric layer. (a) The 3D schematic of our IMPA (below) and its circuit model (upper). In the 3D schematic, the green and yellow components symbolize the $\lambda/4$ and $\lambda/2$ impedance transformers respectively. The red segment represents the superconducting quantum interference device (SQUID) loop. The blue section corresponds to the upper electrode of the capacitor. The purple section denotes the amorphous silicon dielectric layer of the capacitor. (b) Three different fabrication methods of dielectric layer of the capacitor. The first method has the steps of depositing a dielectric layer and then etching it.\cite{Mutus2014} But this often brings problems like over-etching or under-etching. The second method applies a single-layer resist lift-off technique, which might make the dielectric layer not lifted as expected and leaves irregular residual. The third method employs the double-layer resist lift-off technique and it produces a well-patterned square dielectric layer, resulting in a significantly higher fabrication success rate. }
    \label{fig1}
\end{figure*}

In recent years, a variety of schemes have been proposed to enhance the bandwidth and dynamic range of parametric amplifiers.\cite{Roy2015,Esposito2021,Macklin2015,Mutus2014,Duan2021} One scheme is the Josephson traveling-wave parametric amplifier (JTWPA), which achieves a bandwidth over several gigahertz with a gain exceeding 20 dB, making it suitable for the simultaneous readout of multiple qubits.\cite{Macklin2015} However, this scheme requires the fabrication of thousands of nearly identical Josephson junctions, which significantly reduces the overall fabrication success rate of the device. Moreover, the strict requirements on phase matching make the design, fabrication and calibration of JTWPA difficult.\cite{Esposito2021}

Another scheme is the impedance-engineered Josephson parametric amplifier (IMPA), which focuses on designing the environmental impedance to enhance the coupling of the JPA to the environment and reduce the quality factor of the Josephson oscillator, resulting in improving the bandwidth and saturation power.\cite{Roy2015,Mutus2014,Duan2021} In one method, A lumped-element Josephson parametric amplifier was introduced, designed to achieve strong environmental coupling via a tapered transmission line.\cite{Mutus2014} This method successfully broadens the bandwidth and saturation power. However, engineering a tapered transmission line requires complex nano-fabrication. An alternative method of introducing one $\lambda/2$ impedance transformer and one  $\lambda/4$ impedance transformer simplify the fabrication process and also achieves noteworthy performance.\cite{Roy2015,Duan2021} We adopt this method\cite{Duan2021} and present a double-layer resist lift-off process to fabricate the capacitor dielectric layer of IMPA, which offers a higher success rate than traditional methods.

Figure~\ref{fig1a} presents the 3D scheme and the circuit model of our device. The non-linear LC resonator is formed by a capacitor $C_p$ and a superconducting quantum interference device (SQUID) loop as a nonlinear inductor $L_J$. The performance of the JPA is inherently limited by the weak coupling that exists between this nonlinear resonator and its external environment. To enhance the amplifier's gain, Josephson nonlinear components and a coplanar-waveguide (CPW) transformer are integrated on the same chip,\cite{Duan2021} as depicted in Figure~\ref{fig1a}. To begin with, a $\lambda/4$ impedance transformer is employed to convert the standard ambient impedance from $50 \Omega$ to $30 \Omega$. This transformation not only broadens the bandwidth but also reduces the value of $Z_{aux}$. Subsequently, A $\lambda/2$ co-planar waveguide resonator is introduced to induce a positive linear slope in the imaginary component of the impedance shunting the JPA, while maintaining the impedance at $30 \Omega$. This configuration modifies the coherent cavity dynamics, reducing the susceptibility to the effects of detuning the signal from the pump. In our design, the  capacitor $C_p = 3 pF$ and the total critical current ${I_c} = 11.1\mu A$. 

\begin{table*}[]
   \centering
    \begin{tabular}{| p{8.5cm} | p{7cm} |}
    \hline
        Equipment & Device model \\
    \hline
    Maskless Lithography System &Heidelberg Instruments DWL66+ \\
    \hline
    Plasma Enhanced Chemical Vapor Deposition System&Oxford Instruments System100 PECVD\\
    \hline
    Inductively Coupled Plasma Reactive Ion Ecthing & Oxford Instruments Plasmalab System100\\
    \hline
    Electron-Beam Lithography System&JOEL JBX-6300FS\\
    \hline
    Electron-Beam Evaporator System&Plassys MEB550SL3\\
    \hline
\end{tabular}
    \caption{Equipment utilized in the fabrication process.}
    \label{table1}
\end{table*}

Our device is fabricated on a 430-$\mu m$-thick, 2-inch diameter sapphire wafer with a (0001) orientation. The wafer undergoes thermal annealing at 200°C for 3 hours under high vacuum conditions and subsequently subjected to electron beam evaporation (EBD) to deposit a 100-nm-thick aluminum film. Then we pattern the aluminum film with laser direct writing and wet etching techniques, which defines the capacitor's ground electrode and co-planar waveguide resonators. Subsequently, we proceed with the preparation of the dielectric layer of the capacitor. Initially, we adopt the method including a deposition of an amorphous silicon layer and then pattern it by etching,\cite{Mutus2014} illustrated in figure~\ref{fig1b} as the first method. However, this method necessitates precise control of the etching rate of reactive ion etching (RIE), as improper control can lead to issues such as over-etching and under-etching. Over-etching can potentially induce damage to the aluminum film surface, thereby increasing electron scattering and thermal noise, which in turn reduces the signal-to-noise ratio. Under-etching, on the other hand, can leave residual dielectric material, which may interfere with the subsequent growth of SQUIDs. To address these challenges, we investigate alternative methods utilizing lift-off techniques. We finally choose the third method in figure~\ref{fig1b}, a double-layer resist lift-off technique combined with plasma-enhanced chemical vapor deposition (PECVD), to fabracite 250-nm-thick amorphous silicon dielectric layer. Our findings reveal that the single-layer resist lift-off process is prone to defects, such as incomplete removal of the dielectric layer, which can adversely affect the electrical connection between the upper electrode and other components. Conversely, the double-layer resist lift-off process effectively substantially minimizes the accumulation of dielectric residues, thus leading to enhanced process reliability. Furthermore, the potential issue of protruding dielectric layer edges following the lift-off process is considered, as this could interfere with deposition on the upper electrode of the capacitor, thereby impeding proper electrical connection on either side of the protrusion. To address this concern, LOR3A and SPR955 are carefully selected from a range of photoresists, and their spin-coating speeds are precisely controlled to regulate the height of these protrusions. This ensures that the thickness of the aluminum film deposited on the upper electrode exceeds the height of the protrusions, which, results in preserving electrical continuity and improving the success rate of fabrication.

Subsequently, the upper electrode of the capacitor, consisting of a 300-nm-thick aluminum film, is also fabricated using the aforementioned double-layer resist lift-off technique. Ultimately, SQUIDs are fabricated using Dolan bridge shadow evaporation technique.\cite{Dolan1977} A double-layer resist system is employed for the bridge fabrication, consisting of a methyl methacrylate (MMA) EL9 copolymer resist as the first layer and polymethyl methacrylate (PMMA) 950 A5 as the second layer. An in-situ argon ion milling process is essential for optimal electrical connectivity between the initial alumium layer and the SQUIDs. After the ion milling process, We utilize a standard double-angle evaporation technique to create $Al/Al_2O_3/Al$ trilayer structures as Josephson junctions. Several additional SQUIDs are integrated on the chip to verify that the resistance of the junction at ambient temperature is adequate. The types and specific models of instruments utilized in the fabrication process are presented in Table 1. 

Our IMPA device is electrically interconnected to an aluminum sample holder through wire bonding, which is subsequently positioned within a magnetically shielded enclosure. The device is cooled to a 15mK base temperature inside a dilution refrigerator and worked in a reflection mode. A very weak signal ($\sim$ -108 dBm to -132 dBm) is input into the measurement system, subsequently amplified and reflected. A 4-8 GHz cryogenic circulator, connected to the IMPA, is utilized to separate the reflected output signal and the input signal. The output signal undergoes further amplification through a commercial high-electron-mobility transistor (HEMT) amplifier operating at 4 K, followed by additional amplification at room temperature before digitization. 

\begin{figure*}[t]
        \centering
        \subfloat[\label{fig2a}]{\includegraphics[width=0.304\textwidth]{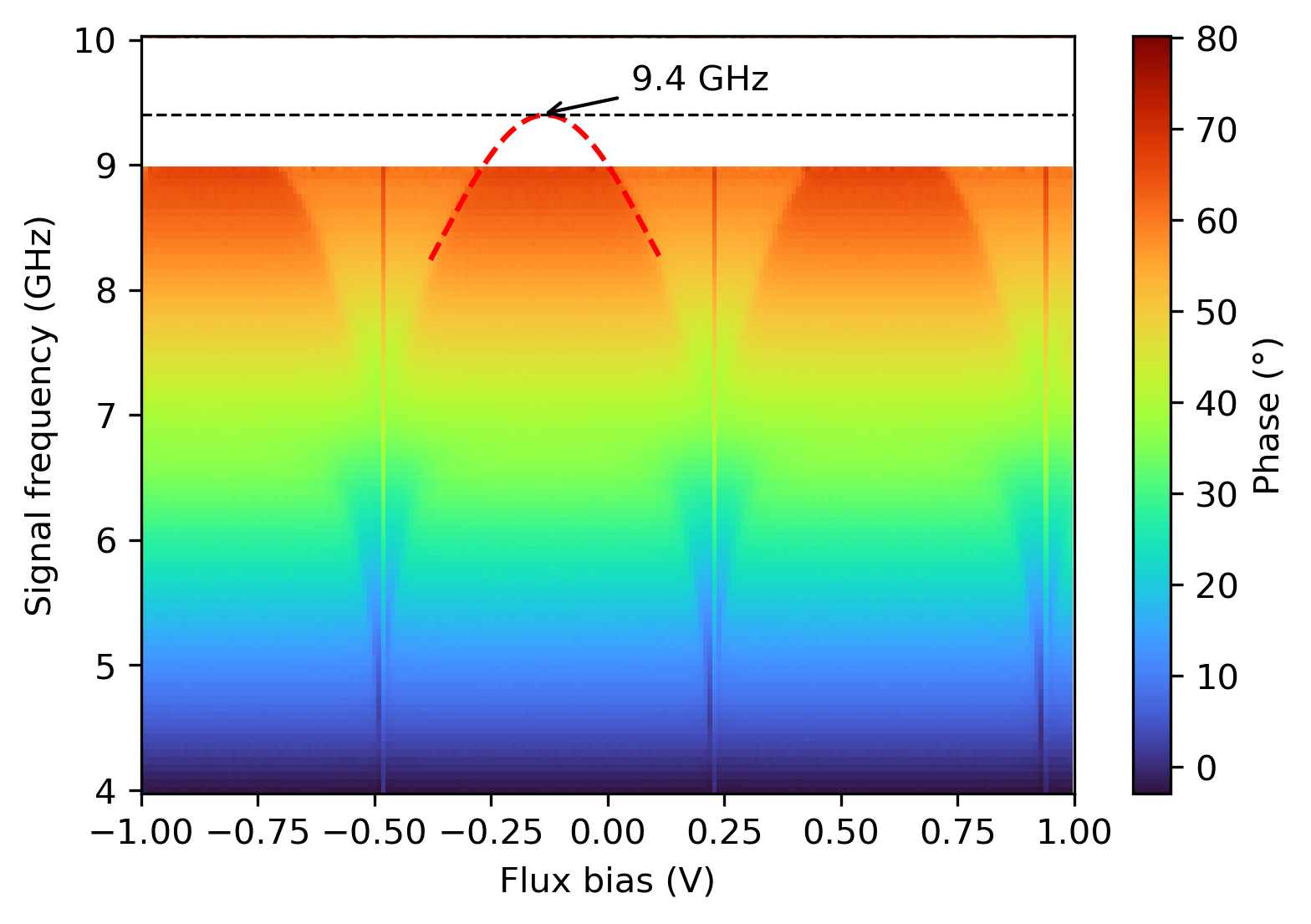}}
        \hfill
        \subfloat[\label{fig2b}]{\includegraphics[width=0.355\textwidth]{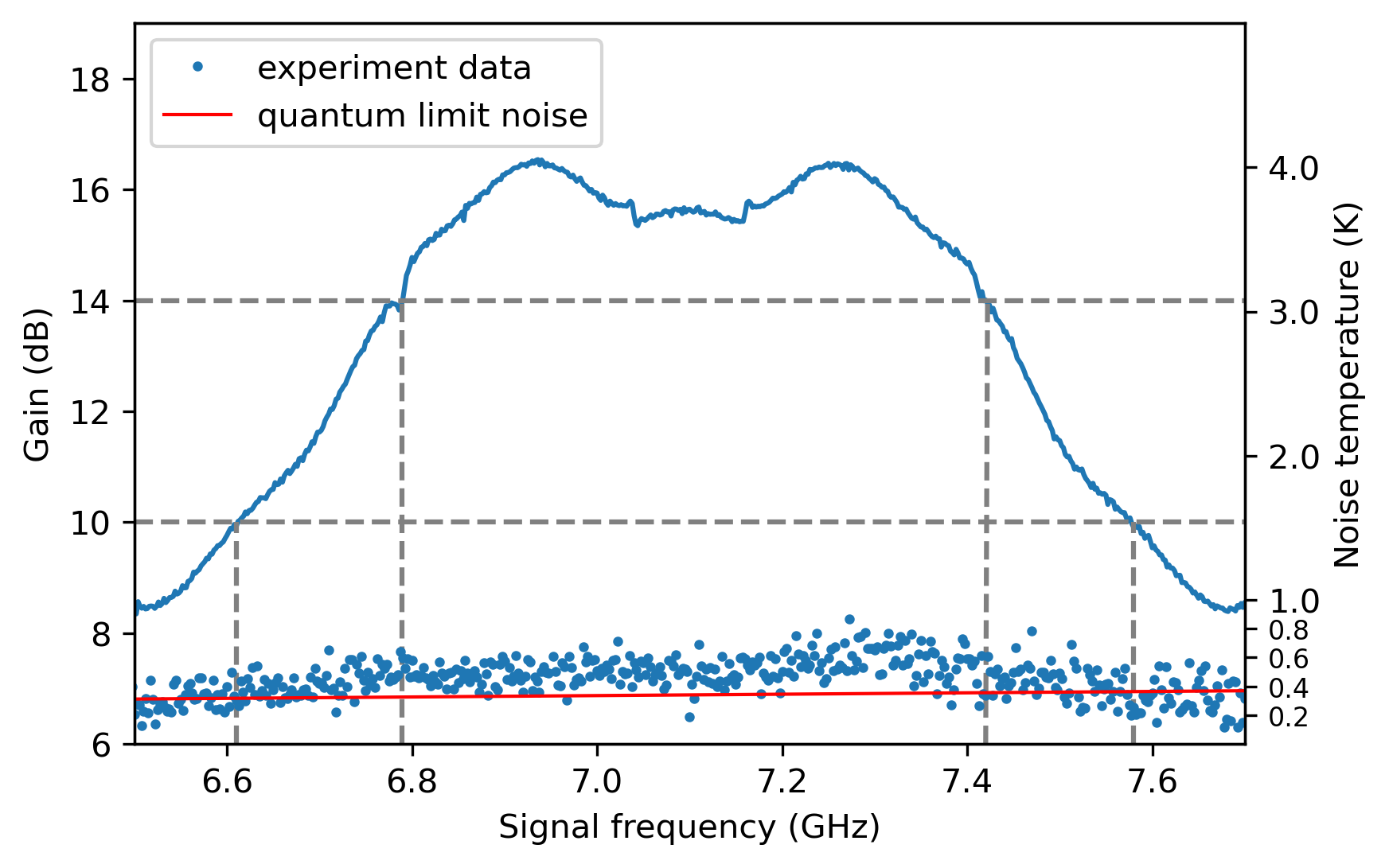}}
        \hfill
        \subfloat[\label{fig2c}]{\includegraphics[width=0.32\textwidth]{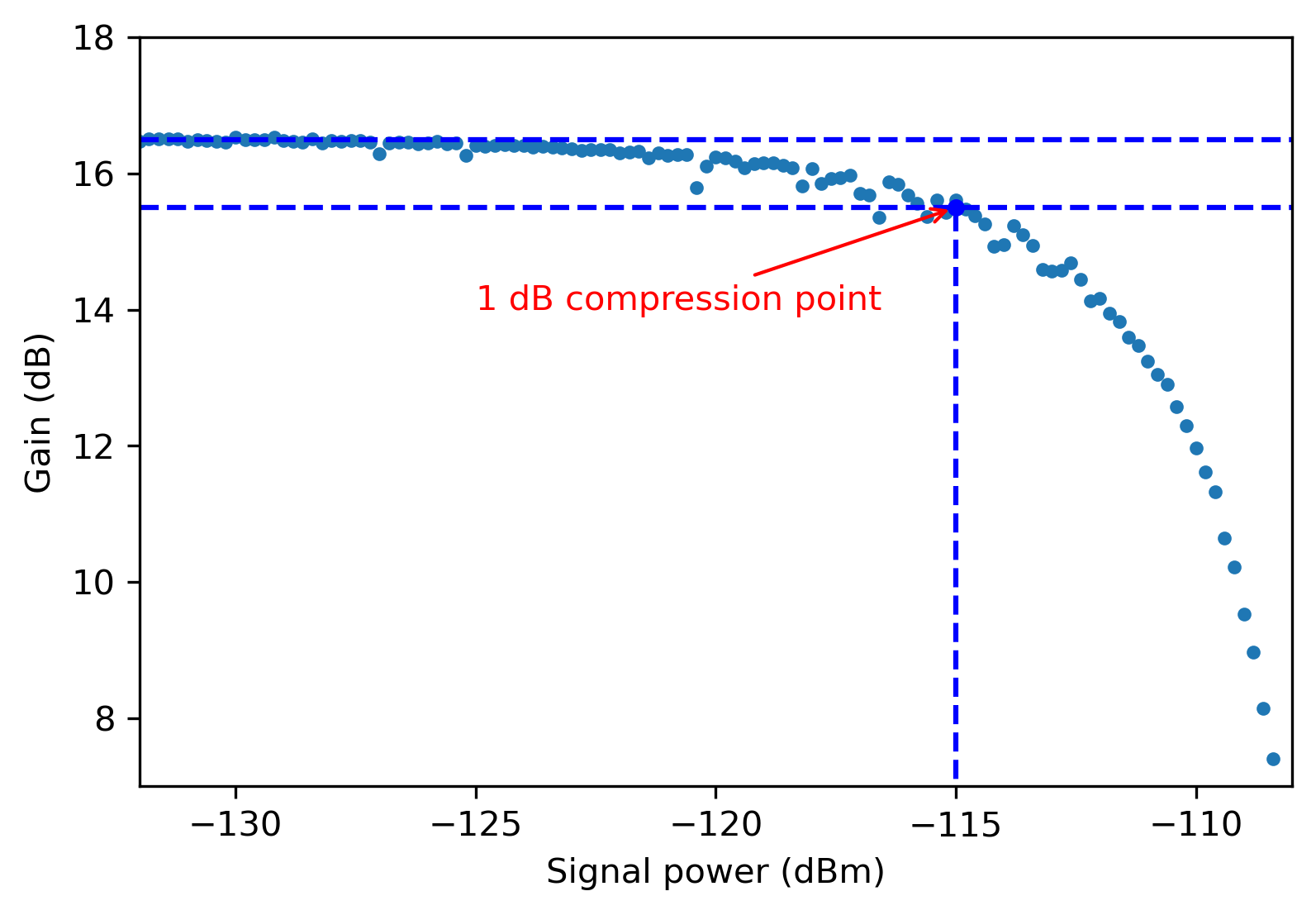}}
        \hfill
        \captionsetup{justification=raggedright,singlelinecheck=false}
    \caption{Tuning, gain and noise performance. (a) The phase component of $S_{21}$ parameter versus the DC flux bias for different signal frequency. A red modulation curve of the phase component is plotted through data fitting, and the peak occurs at a signal frequency of 9.4 GHz. (b) Signal frequency dependence of the gain and noise temperature. The quantum limit noise is defined by the equivalent temperature of one photo energy, symbolized by the solid red line and we employ approximate analytical expressions to characterize the system noise $T_{sys} \approx \frac{Y-1}{G_pG_i^2G_a}T_H$.\cite{Song2021} (c) Gain versus signal power . The saturation power is defined by the 1-dB compression point.}
    \label{fig:group_of_figures}
\end{figure*}

The initial step in characterizing our device is measuring the phase and amplitude components of $S_{21}$ parameter as a function of DC flux bias for various input signal frequency. Typically, the amplitude component of $S_{21}$ exhibits minimal variation with the DC bias due to the low quality factor, whereas the phase component of $S_{21}$ displays more pronounced changes. By varying the DC bias, we observe a periodic structure as illustrated in fig\ref{fig2a}, which is attributed to the modulation of the critical current of the SQUID. Next, we determine the appropriate resonant frequency for signal amplification. We then measure the $S_{21}$ parameter with the IMPA pump on and off and define gain as $\frac{|S_{21}\ pump \ on|}{|S_{21}\ pump\ off|} $. As shown in figure~\ref{fig2b}, the measured gain exhibits a distinct bimodal frequency response. Specifically, the peak gain achieved is 16.5 dB. The bandwidth over which the gain exceeds 10 dB is measured to be approximately 1 GHz, while the bandwidth for which the gain remains above 14 dB is greater than 600 MHz. Figure~\ref{fig2b} also shows the frequency-dependent behavior of noise, with the measured noise temperature of the IMPA closely approaching the quantum limit. Additionally, we investigate the gain as a function of input signal power, as illustrated in figure~\ref{fig2c}, and find that the saturation power of the IMPA is -115 dBm. 

\begin{figure*}
\centering
    \subfloat[\label{fig3a}]{\includegraphics[width=0.485\textwidth, trim=0cm 0cm 0cm 0cm, clip]{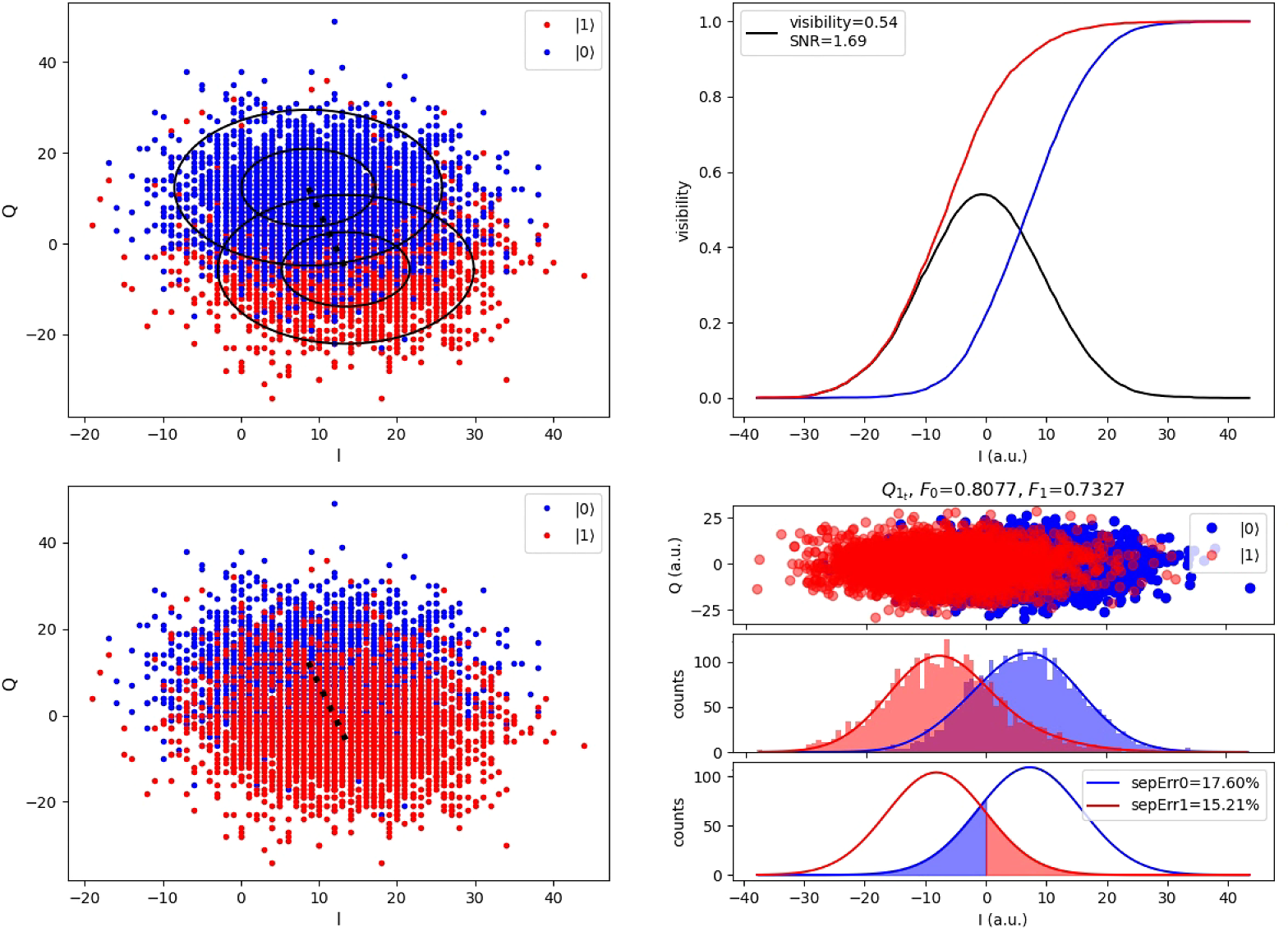}}
    \hfill
    \subfloat[\label{fig3b}]{\includegraphics[width=0.485\textwidth, trim=0cm 0cm 0cm 0cm, clip]{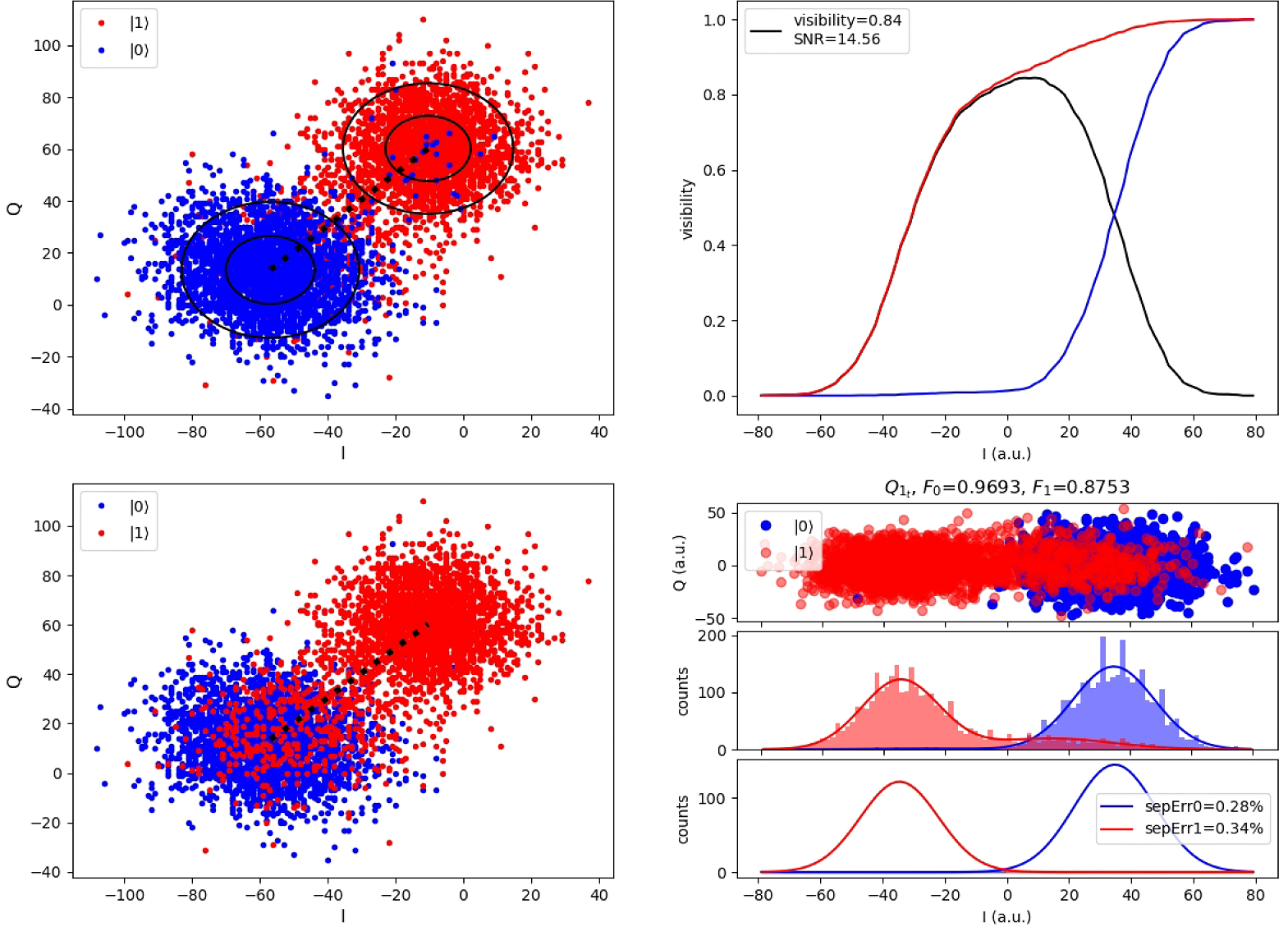}}
    \hfill
    \captionsetup{justification=raggedright,singlelinecheck=false}
\caption{(a) I-Q clouds for qubit states measured with IMPA pump off. (b) I-Q clouds for qubit states measured with IMPA pump on. The I-Q clouds represent the signal scatter due to noise. The dashed lines connecting the center represent the projection axes. The use of the IMPA significantly improves the signal-to-noise ratio.}
    \label{fig3}
\end{figure*}

We fabricate several IMPA devices with identical configurations, utilizing one of them to enhance the measurement quantum efficiency for photon quadrature measurement and tomography.\cite{li2025} In the experiment, which involves the emission of on-demand shaped photons, we further assess the performance of IMPA. The Jaynes-Cummings (JC) model describes a system comprising a two-level atom (such as a superconducting qubit) coupled to a resonator. Denoting $\omega_q$ and $\omega_r$ as the frequencies of the qubit and the resonator, respectively, and $\sigma_z$ as the qubit state operator, the Hamiltonian of the system is given by $H = \frac{1}{2}\hbar\omega_q\sigma_z + \hbar(\omega_r + \chi\sigma_z)\alpha^{\dagger}\alpha$, where the frequency detuning significantly exceeds the coupling rate $g$.\cite{Koch2014} It becomes apparent that the state of the qubit dictates the frequency of the resonator. We then demodulate the signal into its in-phase (I) and quadrature (Q) components. Figure~\ref{fig3a} and \ref{fig3b} display data acquired with the IMPA pump off and on, respectively. The blue and red points represent the readout states $\ket{0}$ and $\ket{1}$. The single-shot measurements are repeated 3,000 times. Notably, with the IMPA activated, the state-discrimination visibility rises from 0.54 to 0.84 and the separation between the two clouds of data points becomes more distinct. This enhanced separation facilitates faster readout processes\cite{Jeffrey2014} and improves the measurement fidelity. 

In the absence of IMPA pump and bias, we measure the relaxation time $T_1$, the coherence time $T_2$ and the signal-to-noise ratio (SNR) to be 5.151 $\mu$s, 3.007 $\mu$s and 1.69, respectively. In contrast, when the IMPA pump and bias are applied, $T_1$ increases to 5.539 $\mu$s, $T_2$ rises to 3.036 $\mu$s and the SNR improves significantly to 14.56. These results demonstrate that the pump and bias of our IMPA significantly enhance the signal-to-noise ratio, while the impact of pump and bias on the decoherence properties of the qubits remains negligible.
In the experimental setup for heterodyne measurement of quantum microwave field, we employ the IMPA to enhance SNR, thereby improving the stability of the quantum state tomography (QST) data. As a result, for the $1/{\sqrt{2}}(\ket{0}+\ket{1})$ quantum state, the coefficient of variation of the QST data improves from 6.73\% with IMPA off to 0.91\% with IMPA on after $3\times10^6$ repetitions. Moreover, after $3\times10^7$ repetitions, the coefficient of variation of QST data improves from 3.22\% with IMPA off to 0.19\% with IMPA on. These results demonstrate the excellent performance in SNR enhancement in real experimental applications. 

The overall quantum efficiency $\eta$ accounts for contributions from noise introduced by the amplifiers, losses and spurious reflections throughout the amplification chain, as well as inefficiencies during the detection process, which includes the sampling and data processing stages.~\cite{li2025} Given the significant gain of the IMPA, the system's noise is primarily dominated by the noise introduced by the first near-quantum-limited amplifier.~\cite{Blais2021} We calibrate the average photon number of the thermal noise induced by the amplification chain as $n_{noise} = \langle \hat{h}^{\dagger} \hat{h} \rangle = 2.78$, which yields an overall quantum efficiency of $\eta = \frac{1}{1 + n_{noise}} \approx 0.26$. This quantum efficiency surpasses the value reported by Ferreira et al.~\cite{Ferreira2024} in a similar setup using a single Josephson traveling-wave parametric amplifier (JTWPA), where they measured $\eta \approx 0.22$. This comparison highlights the exceptional performance of the IMPA from a certain perspective.

We have fabricated and characterized several broadband impedance-transformed Josephson parametric amplifiers (IMPAs). One method has been introduced for preparing the capacitor's dielectric layer, employing a double-layer lift-off process that offers improved stability and accelerates production compared with the traditional technique. Experimental results indicate that the IMPA achieves an instantaneous bandwidth over 950 (600) MHz with a gain exceeding 10 (14) dB, along with saturation input power of -115 dBm and near quantum-limited noise. Due to no significant degradation of the relaxation time and coherence time of the superconducting qubits, the backaction from the IMPA on qubits can be ingored. The IMPA significantly improves the SNR and enables the amplification chain to achieve a high quantum efficiency with $\eta \approx 0.26$, making it a critical necessity for large-scale quantum computation.

This work was supported by the National Natural Science Foundation of China (Grant Nos. 12204528, 92265207, T2121001, 92365301, T2322030, 12122504, 12274142, 12475017), the Innovation Program for Quantum Science and Technology (Grant No. 2021ZD0301800), the Beijing Nova Program (Grant No. 20220484121), the Natural Science Foundation of Guangdong Province (Grant No. 2024A1515010398). This work was supported by the Synergetic Extreme Condition User Facility (SECUF). Devices were made at the Nanofabrication Facilities at Institute of Physics, CAS in Beijing.


%
%

%


\section*{AUTHOR DECLARATIONS}
\subsection*{Conflict of Interest}
The authors have no conflicts to disclose.

\subsection*{Author Contributions}
Zhengyang Mei and Xiaohui Song contributed equally to this work.
\begin{flushleft}
\textbf{Zhengyang Mei}: Conceptualization (equal); Data curation (equal); Formal 
analysis (equal); Investigation (lead); Methodology (supporting); Validation 
(equal); Visualization (lead); Writing – original draft (equal). \textbf{Xiaohui Song}: Conceptualization (equal); Data curation (equal); Formal analysis (equal); Investigation (supporting); Methodology (equal); Project administration (lead); Visualization (supporting); Writing – original draft (equal). \textbf{Xueyi Guo}: Data curation (supporting); Formal analysis (equal); Investigation (supporting); Software (lead). \textbf{Xiang Li}: Data curation (equal); Formal analysis (equal); Investigation (supporting); Software (supporting); Writing – original draft (supporting). \textbf{Yunhao Shi}: Software (supporting); Writing – original draft (supporting). \textbf{Guihan Liang}: Investigation (supporting). \textbf{Chenglin Deng}: Software (supporting). \textbf{Li Li}: Software (supporting). \textbf{Yang He}: Software (supporting). \textbf{Dongning Zheng}: Funding acquisition (equal); Project administration (supporting); Resources (supporting); Supervision(supporting); Writing – review \& editing(supporting). \textbf{Kai Xu}: Funding acquisition (equal); Project administration (supporting); Resources (equal); Supervision(supporting); Writing – review \& editing(supporting). \textbf{Heng Fan}: Funding acquisition (equal); Project administration (supporting); Resources (equal); Supervision(equal); Writing – review \& editing(equal). \textbf{Zhongcheng Xiang}: Conceptualization (equal); Methodology (equal); Funding acquisition (equal); Project administration (supporting); Resources (equal); Supervision(equal); Writing – review \& editing(equal). 
\end{flushleft}

\subsection*{DATA AVAILABILITY}
The data that support the findings of this study are available from the corresponding authors upon reasonable request.

\bibliography{article}

\end{document}